\documentclass{aa}  

\usepackage{graphicx}
\usepackage{txfonts}
\def\XMM{XMM-{\sl Newton}}
\def\swift{{\sl Swift}}
\def\chan{{\sl Chandra}}
\def\beppo{{\sl Beppo}-SAX}

\begin{document}

   \title{Discovery of a fast, broad, transient outflow in NGC 985}

   \author{J. Ebrero\inst{1}
          \and
          G. A. Kriss\inst{2,3}
          \and
          J. S. Kaastra\inst{4,5}
          \and
          J. C. Ely\inst{2}
          }

   \institute{XMM-Newton Science Operations Centre, ESAC,
              Camino Bajo del Castillo s/n, Urb. Villafranca del Castillo, 28692 Villanueva de la Ca\~nada, Madrid, Spain\\
              \email{jebrero@sciops.esa.int}
              \and
              Space Telescope Science Institute, 3700 San Martin Drive, Baltimore, MD 21218, USA
              \and
              Department of Physics and Astronomy, The Johns Hopkins University, Baltimore, MD 21218, USA
              \and
              SRON Netherlands Institute for Space Research, Sorbonnelaan 2, 3584 CA, Utrecht, The Netherlands
              \and
              Leiden Observatory, Leiden University, P. O. Box 9513, 2300 RA, Leiden, The Netherlands
              }

   \date{Received <date>; accepted <date>}

 
  \abstract
   {}
   {We observed the Seyfert 1 galaxy NGC 985 on multiple occasions to search for variability in its UV and X-ray absorption features in order to establish their location and physical properties.}
   {We use \XMM{} to obtain X-ray spectra using the EPIC-pn camera, and the {\it Cosmic Origins Spectrograph} (COS) on the {\it Hubble Space Telescope} (HST) to obtain UV spectra. Our observations are simultaneous and span timescales of days to years.}
   {We find that soft X-ray obscuration that absorbed the low energy continuum of NGC 985 in August 2013 diminished greatly by January 2015. The total X-ray column density decreased from $2.1 \times 10^{22}~\rm cm^{-2}$ to $\sim 6 \times 10^{21} ~\rm cm^{-2}$. We also detect broad, fast UV absorption lines in COS spectra obtained during the 2013 obscuration event. Lines of \ion{C}{iii}*, Ly$\alpha$, \ion{Si}{iv} and \ion{C}{iv} with outflow velocities of $-5970~\rm km~s^{-1}$ and a full-width at half-maximum of $1420~\rm km~s^{-1}$ are prominent in the 2013 spectrum, but have disappeared in all but Ly$\alpha$ in the 2015 spectra. The ionization state and the column density of the UV absorbing gas is compatible with arising in the same gas as that causing the X-ray obscuration. The high velocity of the UV-absorbing gas suggests that the X-ray obscurer and the associated UV outflow are manifestations of an accretion disk wind.}
   {}


   \authorrunning{J. Ebrero et al.}
   \titlerunning{Discovery of a fast, broad, transient outflow in NGC 985}

   \maketitle
%

\section{Introduction}
\label{intro}

X-ray emission is a characteristic footprint of active galactic nuclei (AGN), and it is known to vary at short and long timescales. In some cases, the origin of X-ray variability in AGN could be attributed to the passage of intervening material across our line of sight. This situation is of particular interest as it can be useful to probe the dynamics and physical properties of the matter in the vicinity of the central supermassive black hole (SMBH). Such situations have been reported for the case of short eclipses, possibly due to the passing of broad line region clouds (NGC 1365, \citealt{Ris07}; Mrk 766, \citealt{Ris11}), and for longer periods, as in NGC 5548, where a long-lived stream of matter is obscuring the central source for several years (\citealt{Kaa14}). In fact, the systematic study of \citet{Mar14} shows that this kind of obscuration event may happen recurrently in a majority of AGN but they may pass unnoticed due to the lack of continuous monitoring.

NGC 985 (also known as Mrk 1048), may be another case of a recorded eclipsing event. It is a nearby $z = 0.043$~(\citealt{Fis95}) Seyfert 1 galaxy which shows a distinctive ring-shaped structure, testimony of a recent merger event. NGC 985 also shows a persistent outflow in both the X-ray and the UV, revealed by a multi-component warm absorber (WA) in the X-ray, and blue-shifted absorption troughs in both wavebands (\citealt{Arav02a}; \citealt{Kro05,Kro09}). Following a \swift{} monitoring programme, NGC 985 was discovered to be in a low soft X-ray flux state in July 2013, which triggered a joint \XMM{} and HST observation (\citealt{Par14}). We recently observed NGC 985 using both \XMM{} and the Hubble Space Telescope (HST) in an effort to monitor the time variability of the persistent absorbers and thereby determine their density, distance, mass flux, and kinetic luminosity. Our observations in January 2015 show that the source is back to an overall high X-ray flux state, and therefore emerging from an absorption event. This work is organized as follows: in Sect.~\ref{observations} we describe the observations used and the data reduction methods, in Sect.~\ref{sed} we derive the spectral energy distribution of NGC 985, in Sect.~\ref{results} we report the data analysis, and in Sect.~\ref{discussion} we discuss the results and summarize our findings. We adopt a cosmological framework  with $H_{\rm 0} = 70$~km s$^{-1}$~Mpc$^{-1}$, $\Omega_{\rm M} = 0.3$, and $\Omega_{\Lambda} = 0.7$. The quoted errors refer to 68.3\% confidence level ($\Delta \chi^2 = 1$~for one parameter of interest) unless otherwise stated.


\section{Observations}
\label{observations}

NGC 985 has been observed by \XMM{} in five occasions: once in 2003, and twice in 2013 and 2015. The first observation of 2013 (ObsID 0690870101) was too short ($\sim$20~ks of exposure time) and was excluded from this analysis. The rest of the observations, summarized in Table~\ref{xmmlog}, were retrieved from the \XMM{}~Science Archive (XSA\footnote{{\tt http://xmm.esac.esa.int/xsa/}}) and re-processed using SAS v14.0. The spectra taken with EPIC-pn, which was operated in Small Window mode to prevent pile-up, were extracted using the {\tt xmmselect} task. The source extraction region was a 40 arcsec radius circle centered at the boresight coordinates; the background was extracted using a region of the same size positioned in an area of the detector devoid of X-ray sources. After filtering from background flaring, the EPIC-pn net exposure times were those listed in Table~\ref{xmmlog}.

We observed NGC 985 with the Cosmic Origins Spectrograph (COS) on the Hubble Space Telescope (HST) in two visits coordinated with the most recent \XMM{} observations. We also retrieved a previous COS observation from the Mikulski Archive for Space Telescopes (MAST\footnote{{\tt https://archive.stsci.edu}}) which was coordinated with the prior \XMM{} observation of \citet{Par14}. Table \ref{HSTObsTbl} gives the details of the data sets from the multiple observations. The far-ultraviolet (FUV) channel of COS nominally covers a wavelength range of 1150--1775 \AA\ with a resolving power of $\sim$18,000 (\citealt{Gre12}). Our observations in January 2015 as well as the archival observations from 2013 used gratings G130M and G160M to span these wavelengths. At the redshift of NGC 985, prominent lines such as Ly$\alpha$, \ion{N}{v}, \ion{Si}{iv}, \ion{C}{iv}, and \ion{He}{ii} appear in our spectra. More recently commissioned modes of COS extend wavelength coverage to as short as the Lyman limit at 912 \AA. Our 2015 observations used grating G140L and a central wavelength setting of 1280 to also cover 1000--2000 \AA~at a resolving power of $\sim2000$, which allows us to view the Ly$\beta$ and \ion{O}{vi} transitions in NGC 985. All observations used multiple central wavelength settings and focal-plane positions (FP-POS) to allow our merged spectra to span the gaps between the two detector segments and eliminate flat-field features and detector defects. The individual exposures were combined with updated wavelength calibrations, flat-field procedures, and flux calibrations as described by \citet{Kriss11b} and \citet{DeR15}. The wavelength zero point was adjusted by matching the low-ionization interstellar absorption features in the spectrum to the \ion{H}{i} velocity of $v_{LSR}=-7~\rm km~s^{-1}$ (\citealt{Wak11}).

   \begin{table}
     \centering
     \caption[]{\XMM{}~observations log of NGC 985.}
     \label{xmmlog}
     \begin{tabular}{l c c c}
       \hline\hline
       \noalign{\smallskip}
       ObsID & Obs. Date & Exp. Time & Net Exp. Time \\
              &  (yyyy-mm-dd) & (ks) &    (ks) \\
       \noalign{\smallskip}
       \hline
       \noalign{\smallskip}
       0150470601 & 2003-07-16 & 57.9 & 31.7\\
       0690870501 & 2013-08-10 & 103.7 & 71.6  \\
       0743830501 & 2015-01-13 & 138.9 & 94.0 \\
       0743830601 & 2015-01-25 & 122.0 & 72.0 \\
       \noalign{\smallskip}
       \hline
     \end{tabular}
   \end{table}

\begin{table*}
  \centering
  \caption[]{HST/COS observations of NGC 985}
\label{HSTObsTbl}
\begin{tabular}{l c c c c c}
\hline\hline
\noalign{\smallskip}
Proposal ID & Data Set Name & Grating/Tilt  & Date & Start Time (UT) & Exposure (s)\\
\noalign{\smallskip}
\hline
\noalign{\smallskip}
12953 & lbz801010 & G130M/1327 & 2013-08-12 & 13:04:43 & $\phantom{0}$851 \\
12953 & lbz801020 & G130M/1309 & 2013-08-12 & 13:24:31 & $\phantom{0}$850 \\
12953 & lbz801030 & G160M/1577 & 2013-08-12 & 14:41:47 & 1145 \\
12953 & lbz80040q & G160M/1600 & 2013-08-12 & 16:06:00 & 1141 \\
\noalign{\smallskip}
\hline
\noalign{\smallskip}
13821 & lcjm01010 & G130M/1291 & 2015-01-14 & 02:39:32 & $\phantom{0}$600 \\
13821 & lcjm01020 & G130M/1309 & 2015-01-14 & 02:52:52 & $\phantom{0}$600 \\
13821 & lcjm01030 & G130M/1327 & 2015-01-14 & 03:06:12 & 1505 \\
13821 & lcjm01040 & G160M/1577 & 2015-01-14 & 04:21:22 & 1430 \\
13821 & lcjm01050 & G160M/1611 & 2015-01-14 & 05:37:27 & $\phantom{0}$775   \\
13821 & lcjm01060 & G160M/1600 & 2015-01-14 & 05:53:30 & $\phantom{0}$775   \\
13821 & lcjm01070 & G160M/1589 & 2015-01-14 & 06:09:33 & $\phantom{0}$775   \\
13821 & lcjm01080 & G140L/1280 & 2015-01-14 & 07:13:00 & 5161 \\
13821 & lcjm02010 & G130M/1291 & 2015-01-26 & 02:43:45 & $\phantom{0}$600 \\
13821 & lcjm02020 & G130M/1309 & 2015-01-26 & 02:57:05 & $\phantom{0}$600 \\
13821 & lcjm02030 & G130M/1327 & 2015-01-26 & 03:10:25 & 1505 \\
13821 & lcjm02040 & G160M/1577 & 2015-01-26 & 04:25:14 & 1430 \\
13821 & lcjm02050 & G160M/1611 & 2015-01-26 & 05:41:18 & $\phantom{0}$775  \\
13821 & lcjm02060 & G160M/1600 & 2015-01-26 & 05:57:21 & $\phantom{0}$775  \\
13821 & lcjm02070 & G160M/1589 & 2015-01-26 & 06:13:24 & $\phantom{0}$775 \\
13821 & lcjm02080 & G140L/1280 & 2015-01-26 & 07:16:50 & 5161 \\
\noalign{\smallskip}
\hline
\end{tabular}
\end{table*}


\section{Spectral energy distributions}
\label{sed}

\begin{figure}
  \centering
  \hbox{
  \includegraphics[width=6cm,angle=-90]{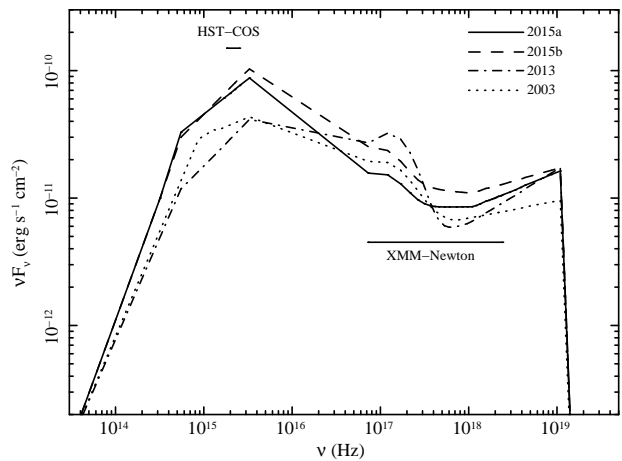}
  }
  \caption{\label{sedfig}Spectral energy distributions of the 2015a (solid line), 2015b (dashed line), 2013 (dot-dashed line), and 2003 (dotted line) observations. The bands covered by HST-COS and \XMM{}~are also indicated.}
\end{figure}

We constructed the spectral energy distributions (SED) of the observations using the contemporaneous measurements of \XMM{}~EPIC-pn in X-rays, HST in the UV (except for the 2003 observation in which no contemporaneous HST data were available), and the \XMM{}~optical monitor (OM) in the UV/optical. The X-ray SED was obtained from the best-fit EPIC-pn model (see Sect.~\ref{xrays}) corrected from instrinsic and Galactic absorption. In the UV we used the HST-COS continuum fluxes at 1175, 1339, 1510, and 1765~\AA~corrected for Galactic extinction using the reddening curve prescription of \citet{Car89}, assuming a color excess $E(B-V)=0.03$~mag based on calculations of \citet{Sch98} as updated by \citet{Sch11}, as given in the NASA/IPAC Extragalactic Database (NED), and a ratio of total to selective extinction $R_{\rm V} \equiv A_{\rm V}/E(B-V)$ fixed to 3.1. The OM observations were taken with the $V$, $B$, $U$, $UVW1$, $UVM2$, and $UVW2$ filters in the 2013 and 2015 observations, except for the 2003 observation, in which only $U$, $UVW1$, $UVM2$, and $UVW2$ filters were used. The OM fluxes were also corrected for Galactic extinction using the reddening curve of \citet{Car89}, including the update in the optical/near-IR of \citet{Don94}. To correct for the host galaxy starlight contribution in the OM filters we used the bulge galaxy template of \citet{Kin96}, scaled to the host galaxy flux at the rest-frame wavelength $F_{\rm gal, 5100 \AA} = 2.48 \times 10^{-15}$~erg cm$^{-2}$~s$^{-1}$~\AA$^{-1}$ (\citealt{Kim08}). The adopted SEDs are shown in Fig.~\ref{sedfig}.

\section{Results}
\label{results}

\subsection{X-rays}
\label{xrays}

The EPIC-pn spectra were analyzed using the {\tt SPEX}\footnote{{\tt http://www.sron.nl/spex}} fitting package (\citealt{Kaa96}) version 2.06.01. The fitting method employed was $\chi^2$ statistics. We adopted a cosmological redshift for NGC 985 of $z = 0.043$~(\citealt{Fis95}). The foreground Galactic column density was set to $N_{\rm H} = 3.17 \times 10^{20}$~cm$^{-2}$~(\citealt{Kal05}) using the {\it hot} model in SPEX, with the temperature fixed to 0.5 eV to mimic a neutral gas. Throughout the analysis we assumed proto-Solar abundances of \citet{LP09}. The spectra were fitted in the $0.3-10$~keV range, and the data were binned using the optimal binning command {\it obin}.

The continuum of the EPIC-pn spectra was modeled in the same way for all the observations. A power-law model provides an unacceptable fit in all the cases: reduced $\chi^2_\nu \sim 25$ for 234 and 236 degrees of freedom in the 2015 observations, $\chi^2_\nu \sim 14$ for 219 degrees of freedom in the 2013 observation, and $\chi^2_\nu \sim 6$ for 221 degrees of freedom in the 2003 observation. There were positive and negative residuals in the soft X-ray band, signatures of intrinsic absorption and a soft excess. The latter was modeled using a modified black body ({\it mbb} model in SPEX), which takes into account modifications of a simple black body model by coherent Compton scattering based on the calculations of \citet{KB89}. The addition of this component improved the fit by $\Delta \chi^2/\Delta \nu = 420/2$, $1875/2$, $3197/2$, and $3734/2$ in the 2003, 2013, and both 2015 observations, respectively. Interestingly, a single power law did not render a good fit of the overall continuum, even if we took into account the effects of intrinsic absorption (see below), showing a persistent hard excess above the Fe K region, particularly significant in the 2015 observations. A broken power-law model was used instead, improving the overall continuum fit and rendering a very significant improvement of the fit of $\Delta \chi^2/\Delta \nu = 168/2$, and $232/2$ in the two 2015 observations, respectively. This effect is less evident, but still significant, in the 2003 and 2013 observations, with $\Delta \chi^2/\Delta \nu = 11/2$, and $8/2$, respectively. The Fe K$\alpha$ line was modeled with a Gaussian line, whose parameters did not vary significantly between observations, thus indicating an origin in matter distant from the central SMBH. In addition, the 2003 pn spectrum shows a small emission line at $\sim 8.8$~keV, which is attributed to a Cu instrumental feature. Its presence was also reported in \citet{Par14}, and it has no impact in the results and conclusions presented in this paper.

\begin{figure}
  \centering
  \hbox{
  \includegraphics[width=7cm,angle=-90]{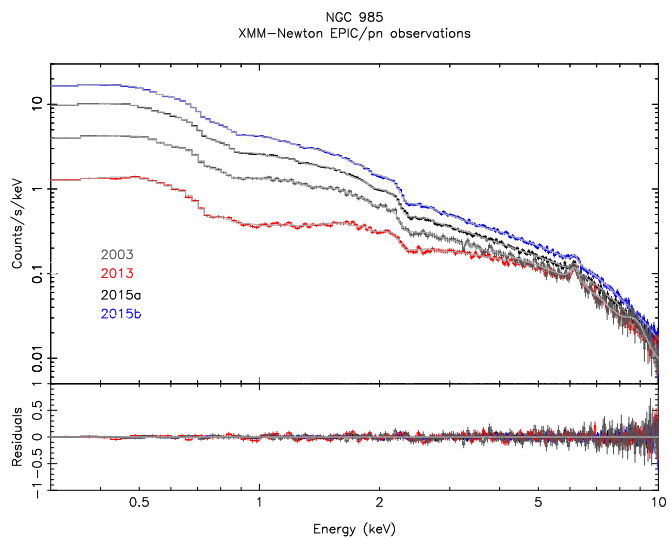}
  }
  \caption{\label{pnfig}\XMM{}~EPIC-pn spectra of NGC 985 in 2003, 2013, and 2015. The solid grey line represents the best-fit model.}
\end{figure}

\begin{table*}
  \centering
  \caption[]{Best-fit results of the \XMM{}~EPIC-pn spectra of NGC 985.}
  \label{xtab}
  \renewcommand\arraystretch{1.1}
  \begin{tabular}{l|l c c c c}
    \hline\hline
    \noalign{\smallskip}
    Model  &  Parameter &  Obs. 2003  &  Obs. 2013  &  Obs. 2015a  &  Obs. 2015b \\
    \noalign{\smallskip}
    \hline
    \noalign{\smallskip}
    Power-law & ${\rm F_{(0.3-10~keV)}}$\tablefootmark{a} & $1.11^{+0.12}_{-0.08} \times 10^{-11}$ & $8.79^{+0.64}_{-0.42} \times 10^{-12}$ & $2.02^{+0.12}_{-0.06} \times 10^{-11}$ & $2.93^{+0.06}_{-0.04} \times 10^{-11}$ \\
             & $\Gamma$\tablefootmark{b} & $2.21^{+0.09}_{-0.03}$ & $1.71^{+0.07}_{-0.03}$ & $2.14 \pm 0.02$ & $2.22 \pm 0.02$ \\
             & $E_{\rm 0}$\tablefootmark{c} &  $3.72^{+0.19}_{-0.96}$  &  $2.69 \pm 0.15$ & $4.54 \pm 0.17$ & $3.54 \pm 0.07$ \\
             & $\Delta\Gamma$\tablefootmark{d} &  $-0.37 \pm 0.07$  & $-0.19^{+0.05}_{-0.07}$ & $-0.42 \pm 0.02$ & $-0.37 \pm 0.02$ \\
    \noalign{\smallskip}
    \hline
    \noalign{\smallskip}
    Mod. BB  & ${\rm F_{(0.3-10~keV)}}$\tablefootmark{a} & $2.7^{+0.8}_{-0.2} \times 10^{-12}$ & $9.3 \pm 1.6 \times 10^{-13}$ & $1.7 \pm 0.1 \times 10^{-12}$ & $2.5 \pm 0.2 \times 10^{-12}$ \\
             & $T$\tablefootmark{e} & $0.232 \pm 0.007$ & $0.211 \pm 0.004$ & $0.160 \pm 0.005$ & $0.146 \pm 0.003$ \\
    \noalign{\smallskip}
    \hline
    \noalign{\smallskip}
    Gauss    & ${\rm F_{line}}$\tablefootmark{a} & $2.4 \pm 0.3 \times 10^{-13}$ & $1.8 \pm 0.2 \times 10^{-13}$ & $1.9 \pm 0.2 \times 10^{-13}$ & $1.5 \pm 0.2 \times 10^{-13}$ \\
             & $E$\tablefootmark{f} & $6.48 \pm 0.02$ & $6.47 \pm 0.02$ & $6.50 \pm 0.01$ & $6.47 \pm 0.02$ \\
             & FWHM\tablefootmark{g} & $0.39 \pm 0.06$ & $0.22 \pm 0.04$ & $0.28 \pm 0.04$ & $0.21 \pm 0.05$ \\
    \noalign{\smallskip}
    \hline
    \noalign{\smallskip}
   {\it xabs} 1  & $\log \xi$\tablefootmark{h} & $2.84 \pm 0.03$ & $2.86 \pm 0.03$ & $2.81 \pm 0.01$ & $2.78 \pm 0.02$ \\
             & $N_{\rm H}$\tablefootmark{i} & $7.2^{+5.2}_{-4.0} \times 10^{22}$ & $2.1 \pm 1.0 \times 10^{22}$ & $9.8^{+2.8}_{-1.8} \times 10^{22}$ & $2.4 \pm 0.3 \times 10^{22}$\\
    \noalign{\smallskip}
    \hline
    \noalign{\smallskip}
   {\it xabs} 2  & $\log \xi$\tablefootmark{h} & $1.88 \pm 0.02$ & $1.83 \pm 0.06$ & $2.03 \pm 0.02$ & $2.18 \pm 0.09$ \\
             & $N_{\rm H}$\tablefootmark{i} & $1.8 \pm 0.2 \times 10^{22}$ & $1.0 \pm 0.3 \times 10^{22}$ & $8.8 \pm 0.8 \times 10^{21}$ & $4.6^{+1.1}_{-0.5} \times 10^{21}$ \\
    \noalign{\smallskip}
    \hline
    \noalign{\smallskip}
    {\it xabs} 3  & $\log \xi$\tablefootmark{h} & $\dots$ & $1.58^{+0.12}_{-0.21}$ & $1.29 \pm 0.12$ & $1.80 \pm 0.05$ \\
             & $N_{\rm H}$\tablefootmark{i} & $\dots$ & $4.0^{+1.4}_{-3.0} \times 10^{21}$ & $1.3 \pm 0.4 \times 10^{21}$ & $4.0^{+0.9}_{-0.3} \times 10^{21}$ \\
    \noalign{\smallskip}
    \hline
    \noalign{\smallskip}
    {\it xabs} 4  & $\log \xi$\tablefootmark{h} & $-1.29 \pm 0.12$ & $-0.94 \pm 0.13$ & $-0.84 \pm 0.24$ & $-0.55^{+0.90}_{-0.30}$ \\
   (Obscurer)    & $N_{\rm H}$\tablefootmark{i} & $1.4^{+0.2}_{-0.5} \times 10^{22}$ & $2.1 \pm 0.2 \times 10^{22}$ & $5.5^{+1.2}_{-0.5} \times 10^{21}$ & $7.0^{+1.2}_{-2.1} \times 10^{21}$ \\
             & $f_{\rm C}$\tablefootmark{j} & $0.85^{+0.02}_{-0.18}$ & $0.92 \pm 0.01$ & $0.30 \pm 0.04$ & $0.18 \pm 0.03$ \\
    \noalign{\smallskip}
    \hline
    \noalign{\smallskip}
    $\chi^2$/d.o.f. &  &  $200/212$ & $262/208$ & $300/223$ & $298/225$ \\
    \noalign{\smallskip}
    \hline
  \end{tabular}
  \tablefoot{
    \tablefoottext{a}{Observed flux, in units of erg cm s$^{-1}$;}
    \tablefoottext{b}{Power-law photon index, corresponds to the slope of the low-energy branch;}
    \tablefoottext{c}{Power-law energy break, in units of keV;}
    \tablefoottext{d}{Power-law photon index break, corresponds to the slope of the high-energy branch of the spectrum when substracted from $\Gamma$;}
    \tablefoottext{e}{Modified black body temperature, in units of keV;}
    \tablefoottext{f}{Line energy, in units of keV;}
    \tablefoottext{g}{Line FWHM, in units of keV;}
    \tablefoottext{h}{Ionization parameter, in units of erg cm s$^{-1}$;}
    \tablefoottext{i}{Hydrogen column density, in units of cm$^{-2}$;}
    \tablefoottext{j}{Covering factor.}
    }
\end{table*}


Absorption caused by ionized gas was modeled using the {\it xabs} model, which calculates the transmission of a slab of material where all the ionic column densities are linked through a photoionization balance model computed with CLOUDY version 13.01 (\citealt{Fer13}), using as inputs the SED of NGC 985 described in Sect.~\ref{sed}. We used as free parameters in this model the hydrogen column density $N_{\rm H}$ and the ionization parameter, defined as $\log \xi = L/nR^2$, where $L$ is the ionizing luminosity in the $1-1000$~Ryd range, $n$ is the density of the gas, and $R$ is the distance to the central ionizing source. The addition of a single {\it xabs} component significantly improved the fit ($\Delta \chi^2/\Delta \nu = 477/2$, $572/2$, $1861/2$, and $2236/2$ in the 2003, 2013, and both 2015 observations, respectively.), although several absorption residuals remained in the soft X-ray range of the spectra. A second WA was therefore added, further improving the fit by $\Delta \chi^2/\Delta \nu = 12/2$, $71/2$, $265/2$, and $310/2$. The presence of some residuals below $\sim$1.5~keV in the 2015 observations motivated the inclusion of an additional WA ($\Delta \chi^2/\Delta \nu = 84/2$, and $63/2$). The 2013 observation allowed also for this extra component ($\Delta \chi^2/\Delta \nu = 77/2$), but its addition to the model in the 2003 observation no longer improved the fit. These WAs have ionization parameters ranging from $\log \xi \sim 1.3$ to $\sim 2.8$, and column densities in the range of $N_{\rm H} \sim 10^{21}$~to a few times $10^{22}$~cm$^{-2}$.

Furthermore, we modeled the transient obscuration seen in the 2013 observation with a neutral absorber, fixing the ionization parameter to $\log \xi \equiv -4$, leaving the column density $N_{\rm H}$ and covering factor as free parameters (\citealt{Par14}). This improved the overall fit by $\Delta \chi^2/\Delta \nu = 131/2$ but it left absorption residuals at $\sim 1$~keV. A much better fit, however, was obtained when we set the ionization parameter free ($\Delta \chi^2/\Delta \nu = 18/1$), indicating that this obscuration is not caused by fully neutral matter but by a mildly ionized gas ($\log \xi = -0.94 \pm 0.13$) with $N_{\rm H} = 2.1 \pm 0.2 \times 10^{22}$ and high covering factor ($0.92 \pm 0.01$). This component was the ultimate cause of the low soft X-ray flux state. Interestingly, the 2015 spectra required some of this additional absorption, $\Delta \chi^2/\Delta \nu = 20/3$, and $15/2$, albeit with a much lower covering factor, in line with the results of the contemporaneous HST-COS observations (see Sect.~\ref{uv}). The 2003 observation also required this additional very low ionization obscurer ($\Delta \chi^2/\Delta \nu = 57/3$), in a somewhat intermediate state between the 2013 and 2015 observations. We note that due to the resolution of EPIC-pn it is very difficult to appropriately constrain all the WA parameters involved in this fit. Some of them, like the outflow velocity of the intervening gas, can only be modeled through the analysis of high-resolution spectra. We stress, however, that the presence of a multi-component WA is significantly required to obtain a good fit to the data. A detailed analysis of the WA in NGC 985 using the RGS spectra of these observations will be reported in a forthcoming paper (Ebrero et al., in preparation). The best-fit results are reported in Table~\ref{xtab}, and the different EPIC-pn spectra are shown in Fig.~\ref{pnfig}.

\subsection{UV}
\label{uv}

Our HST observations of NGC 985 show the usual bright, blue continuum and broad emission lines typical of Seyfert 1 galaxies with an overall appearance similar to all prior UV observations of this object. The six components of the persistent UV absorption enumerated by \citet{Arav02a} are prominent in all three epochs of COS spectra. The most surprising new feature, however, is highly blue-shifted broad absorption that appears for the first time in 2013, and shows some residual traces in Ly$\alpha$ in our 2015 spectra.

We first noticed the potential Ly$\alpha$ absorption feature in our 2015 COS observations. We eliminated the possibilitiy that this was merely an instrumental artifact by examining the individual spectra acquired at different central wavelength settings and focal-plane positions. The feature appeared in all individual spectra in both observations. We also checked the white dwarf spectra acquired each month to monitor the sensitivity variations in COS (\citealt{Hol14}). The Ly$\alpha$ feature does not appear in any of the preceding or following monitor spectra. Finally, we examined the archival spectra of NGC 985 from 2013, where we saw the much stronger Ly$\alpha$ absorption as well as the troughs from several other ions.

\begin{figure*}
  \centering
  \includegraphics[width=14cm, angle=-90]{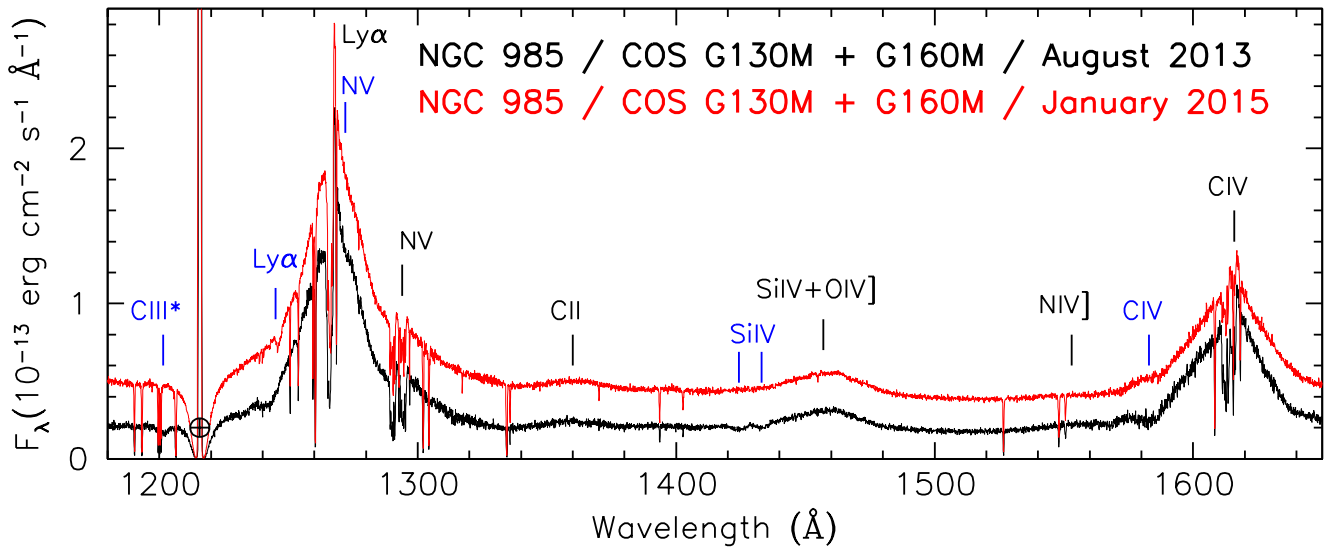}
  \vspace{-6cm}
  \caption{Calibrated and merged COS spectrum of NGC 985 from 2013 (black) and 2015 (red). The spectrum has been binned by 8 pixels (1 resolution element) for display purposes. We indicate the most prominent emission features with black labels and tick marks; blue labels indicate the broad absorption features that are most prominent in the 2013 spectrum. An Earth symbol is atop the geocoronal Ly$\alpha$ emission line in the center of the Milky Way \ion{H}{i} absorption trough.}
  \label{HSTspectra}
\end{figure*}

Figure \ref{HSTspectra} compares the COS spectrum from 2013, when the heavy X-ray obscuration was present, to the average spectrum from our two observations in 2015, when the low-energy X-ray absorption had dissipated. The 2015 spectrum shown in Figure \ref{HSTspectra} is a time-weighted average of the individual exposures obtained in our two visits. NGC 985 varied slightly in flux between the two visits. On 2015-01-14, the continuum flux was $\rm F_\lambda(1410) = 4.1 \times 10^{-14}~ergs~cm^{-2}~s^{-1}~\AA^{-1}$. Twelve days later, on 2015-01-26, it was $\rm F_\lambda(1410) = 4.6 \times 10^{-14}~ergs~cm^{-2}~s^{-1}~\AA^{-1}$. In the 2013 spectrum, one can see that the broad UV absorption was prominent in transitions of \ion{C}{iii}*, Ly$\alpha$, \ion{Si}{iv}, and \ion{C}{iv}, but shows only traces in Ly$\alpha$ in the 2015 spectrum. \ion{N}{v} absorption may be present in 2013, but it is difficult to measure since it falls on top of the peak of the Ly$\alpha$ emission line.

To measure the characteristics of the broad UV absorption lines, we fit a multi-component model to the continuum, emission lines, and broad absorption lines that is similar to our model of NGC 5548 (\citealt{Kaa14}). We use the {\tt specfit} fitting package (\citealt{Kriss94}) to model the spectrum. For the continuum, we use a reddened power law in $f_{\lambda}$ with $\rm E(B - V) = 0.03$ (\citealt{Sch11}; \citealt{Sch98}) absorbed by foreground Galactic \ion{H}{i} with a column density of $3.17 \times 10^{20}~\rm cm^{-2}$ (\citealt{Kal05}). The emission lines are treated as multiple Gaussian emission components, and the broad absorption lines as Gaussian components in negative flux. Since the broad absorption lines are highly blue shifted, the details of the emission line profiles do not affect our measurements, and we will discuss the emission model in more detail in a future publication that discusses the narrow absorption lines. For the broad absorption troughs in the COS 2013 spectrum, free parameters include the equivalent width (EW), the velocities, and the full-width-at-half-maximum (FWHM). For \ion{C}{iv} and  \ion{Si}{iv}, the velocities of the doublets are tied together, and the widths are tied as well. Initially we forced all lines to have the same FWHM. In our final iteration, when the FWHM were permitted to vary freely, they all settled at the same best-fit value. In the 2015 spectrum, since only Ly$\alpha$ was visible, we permitted only its parameters to vary freely. For the other lines, we fixed the velocity and the FWHM at the best-fit values obtained for the 2013 spectrum in order to determine the upper limits (1-$\sigma$ confidence) on the equivalent widths. In Table \ref{AbsTbl} we give the velocities, widths, and equivalent widths of the broad absorption lines.

\begin{table*}
  \centering
  \caption[]{Properties of the Broad UV Absorption Lines in NGC 985}
  \label{AbsTbl}
\begin{tabular}{l c c c c c c}
\hline\hline
\noalign{\smallskip}
{Line} & $\lambda_o$ & {EW} & {Velocity} & {FWHM} & $\rm C_f$ & $\rm N_{ion}$\\
      & (\AA)       & (\AA) & ($\rm km~s^{-1}$) & ($\rm km~s^{-1}$) & & ($10^{14}~\rm cm^{-2}$)\\
\noalign{\smallskip}
\hline
\noalign{\smallskip}
{COS 2013} \\
\noalign{\smallskip}
\hline
\noalign{\smallskip}
\ion{C}{iii}* & 1175.67 & $1.70 \pm 0.06$ & $-6130 \pm 50$ & $1420 \pm 50$ & 0.27 & $>1.9$ \\
Ly$\alpha$ & 1215.67  & $1.30 \pm 0.04$ & $-5970 \pm 20$ & $1420 \pm 50$ & 0.21 & $>2.9$ \\
\ion{Si}{iv} & 1393.76 & $1.03 \pm 0.04$ & $-5950 \pm 20$ & $1420 \pm 50$ & 0.14 & $>2.9$ \\
\ion{Si}{iv} & 1402.77 & $0.96 \pm 0.04$ & $-5950 \pm 20$ & $1420 \pm 50$ & 0.13 & $>2.9$ \\
\ion{C}{iv} & 1548.20  & $1.00 \pm 0.04$ & $-5950 \pm 20$ & $1420 \pm 50$ & 0.13 & $>5.3$ \\
\ion{C}{iv} & 1550.77  & $0.94 \pm 0.04$ & $-5950 \pm 20$ & $1420 \pm 50$ & 0.12 & $>5.3$ \\
\noalign{\smallskip}
\hline
\noalign{\smallskip}
{COS 2015} \\
\noalign{\smallskip}
\hline
\noalign{\smallskip}
\ion{C}{iii}* & 1176.12 & $<0.14$ & $-6130$ & $1420$ & $<0.02$ & $<0.15$ \\
Ly$\alpha$ & 1215.67  & $0.26 \pm 0.04$ & $-5300 \pm 10$ & $350 \pm 50$ & $>0.11$ & $0.55$ \\
\ion{Si}{iv} & 1393.76 & $<0.08$ & $-5950$ & $1420$ & $<0.02$ & $<0.1$ \\
\ion{Si}{iv} & 1402.77 & $<0.08$ & $-5950$ & $1420$ & $<0.02$ & $<0.1$ \\
\ion{C}{iv} & 1548.20  & $<0.08$ & $-5950$ & $1420$ & $<0.02$ & $<0.1$ \\
\ion{C}{iv} & 1550.77  & $<0.08$ & $-5950$ & $1420$ & $<0.02$ & $<0.1$ \\
\noalign{\smallskip}
\hline
\end{tabular}
\end{table*}

\begin{figure}
  \centering
   \includegraphics[width=8.5cm]{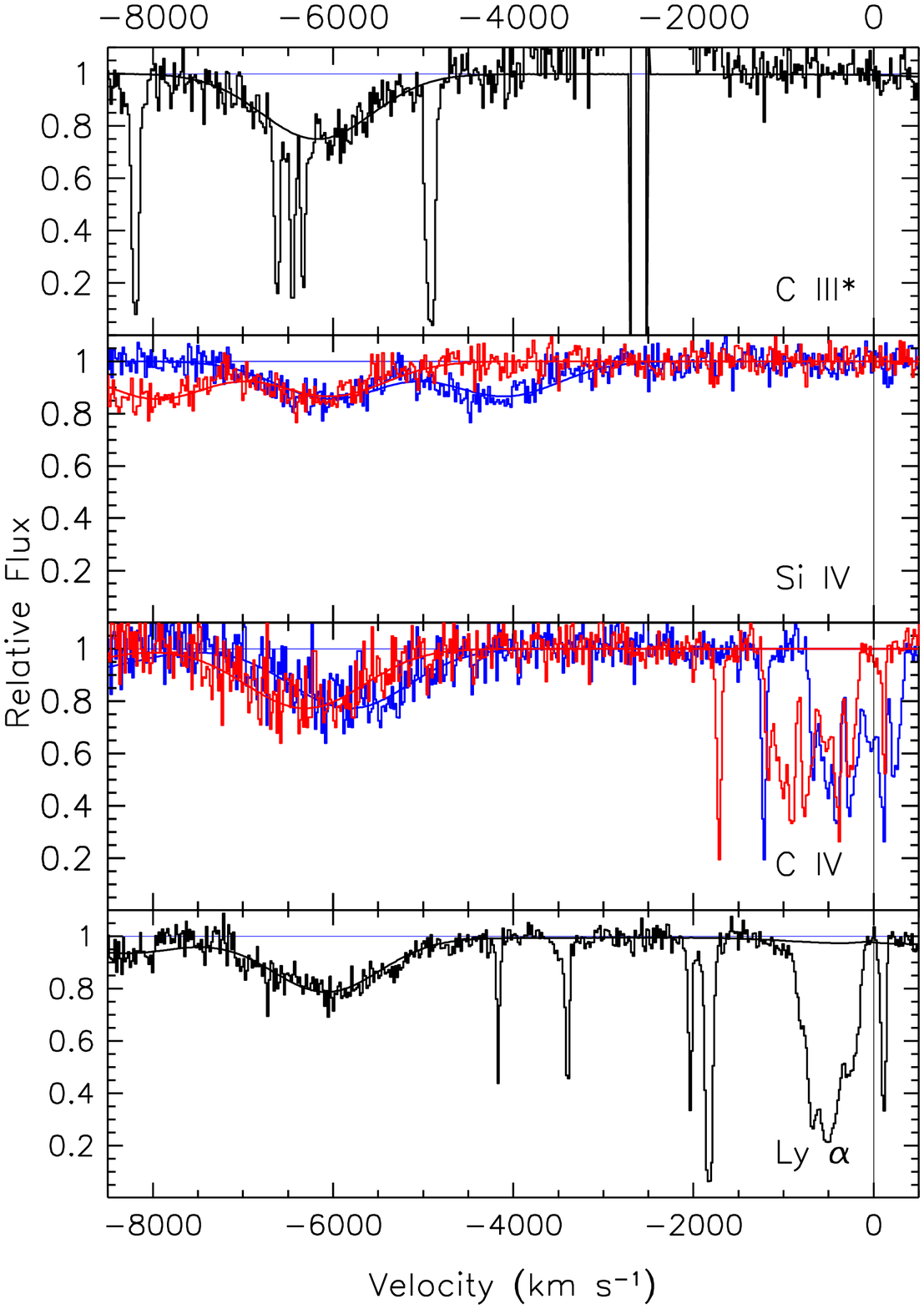}
  \caption{Prominent broad absorption features in the 2013 COS spectrum of NGC 985. We plot normalized relative fluxes as a function of velocity relative to the systemic redshift of $z=0.043$. The spectra have been binned by 8 pixels (1 resolution element) for display purposes. The top panel shows the \ion{C}{iii}* multiplet. The middle two panels show the \ion{Si}{iv} and \ion{C}{iv} absorption troughs with the red component of each doublet as a red line and the blue component as a blue line. The bottom panel shows the Ly$\alpha$ absorption trough. The smooth thin line in each panel shows our fit to the normalized spectrum.}
  \label{AbsNorm}
\end{figure}

Figure \ref{AbsNorm} shows normalized absorption profiles for these absorption lines in the 2013 COS spectrum. From the nearly equal depths of the red and blue components of \ion{Si}{iv} and \ion{C}{iv} as shown by these profiles and the equivalent widths in Table \ref{AbsTbl}, it is likely that these absorption features are saturated, and that they only partially cover the underlying emission regions. The covering fraction $\rm C_f$ given Table \ref{AbsTbl} is determined for coverage of all underlying emission components at the deepest point of each absorption trough. We have insufficient information to unambiguously determine whether the absorption primarily covers only the continuum source, or portions of both the continuum source and the broad emission-line region (BLR), but, given that there is essentially no line emission in the portion of the spectrum absorbed by \ion{C}{iii}*, and since covering fractions for the other ions are lower, it seems likely that the broad absorption we detect primarily covers the continuum emission region and only a small fraction at most of the BLR. In determining the covering fraction, we assume that each absorption line is completely saturated at the center of the trough. To determine the total column densities given in Table \ref{AbsTbl}, we integrated the normalized absorption profiles assuming the troughs represent apparent optical depth (AOD) (e.g., see \citealt{Arav02b}), but, given the high degree of saturation, we simply quote those values as lower limits.

\begin{figure}
  \centering
   \includegraphics[width=7cm, angle=-90]{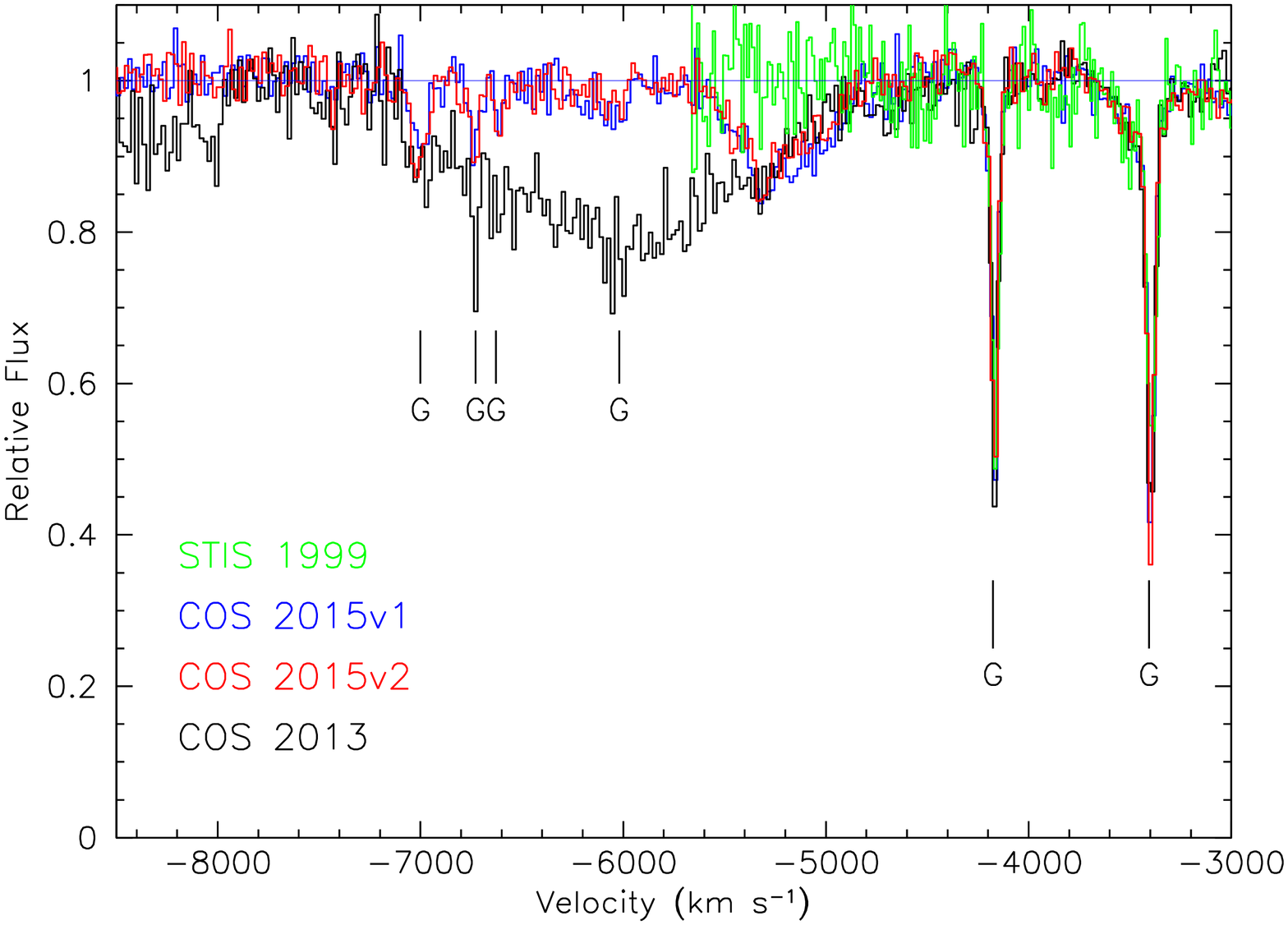}
  \caption{Normalized archival HST spectra of NGC 985 covering the region of Ly$\alpha$ absorption seen in the 2013 and 2015 COS spectra. The 1999 STIS spectrum is in green; the 2013 COS spectrum is in black; the 2015-01-14 COS spectrum is in blue; the 2015-01-26 COS spectrum is in red. For display purposes, the COS spectrum have been binned by 8 pixels (1 resolution element); the STIS spectrum has not been binned. Interstellar absorption features unrelated to NGC 985 are designated with a "G".}
  \label{BroadAbsVary}
\end{figure}

The broad absorption features we detect in the COS 2013 spectrum of NGC 985 vary on approximately the same timescale as the appearance and disappearance of the sof X-ray obscuration reported by \citet{Par14}. Figure \ref{BroadAbsVary} compares the Ly$\alpha$ absorption region in all the HST archival spectra of NGC 985. In the 1999 STIS spectrum analyzed by \citet{Arav02a}, no broad Ly$\alpha$ absorption is present. The broad absorption is prominent in 2013, but it is only weakly present in our two COS spectra obtained in January 2015. For these latter two COS observations, the broad  Ly$\alpha$ absorption has diminished to a weak trough on the red wing of the absorption feature detected in the 2013 spectrum. Our 2015 observations also include short-wavelength G140L spectra that cover the Ly$\beta$ and \ion{O}{vi} regions of the NGC 985 spectrum. While we detect the typical strong \ion{O}{vi} emission line and associated narrow \ion{O}{vi} absorption, we do not see broad \ion{O}{vi} absorption nor Ly$\beta$ absorption. The non-detection of Ly$\beta$ absorption in 2015 is consistent with the Ly$\alpha$ absorption being optically thin in 2015, so in Table \ref{AbsTbl} we quote AOD column densities for the Ly$\alpha$ absorption in the 2015 spectrum, and upper limits for the other ions.

The six narrow absorption components identified by \citet{Arav02a} have full-widths-at-half-maximum of $\sim50~\rm km~s^{-1}$ and span a velocity range of $-810$ to $+210~\rm km~s^{-1}$. The best fit photoionization solution obtained by \citet{Arav02a} gave a total column density of $\rm log~N_H = 21.4$~cm$^{-2}$ for an ionization parameter of $\rm log~U=-0.3$ (equivalent to $\rm log~\xi \sim -1.8$ for the SED given in Sect.~\ref{sed}). In our spectra, these lines have not changed in velocity or width, and they are present only in Ly$\alpha$, Ly$\beta$, \ion{C}{iv}, \ion{N}{v} and \ion{O}{vi}. As in \citet{Arav02a}, the lines are partially saturated, particularly in neutral hydrogen, and they only partially cover the source.  The line depths vary among all four observations. From the variability timescales, we will be able to constrain the density of the absorbing gas in all six components. In concert with photoionization solutions, we will then be able to determine the distance of these absorbing components from the nuclear source. We will present this detailed analysis in a subsequent paper (Ebrero et al., in preparation).


\section{Discussion and conclusions}
\label{discussion}

\begin{figure}
  \centering
  \hbox{
  \includegraphics[width=6.5cm,angle=-90]{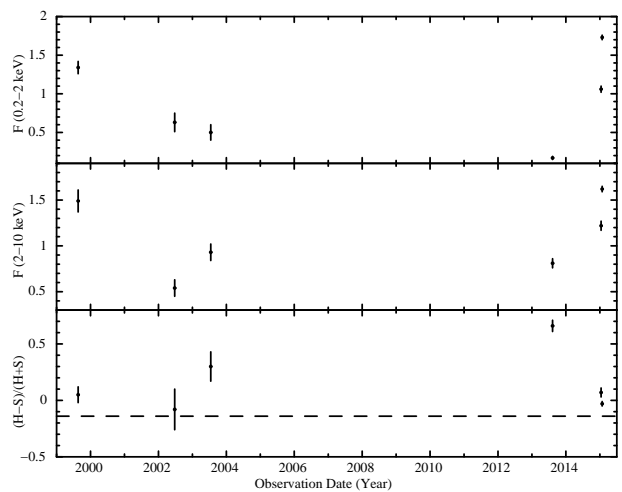}
  }
  \caption{\label{hrfig}Soft ($0.2-2$~keV) and hard ($2-10$~keV) X-ray fluxes (top and central panels, respectively), and hardness ratio (bottom panel) of NGC 985 as measured by \beppo{} (1999), \chan{} (2002), and \XMM{} (2003, 2013, and 2015). Fluxes are in units of $10^{-11}$~erg cm$^{-2}$ s$^{-1}$. The dashed line represents the average hardness ratio from \swift{} observations in 2007-2008, when the source was unobscured.}
\end{figure}

NGC 985 seems to have undergone an eclipsing event caused by mildly ionized gas crossing our line of sight. The main effect of this absorption is the decrease in the soft X-ray flux while the hard X-rays are mostly unaffected, which leads to a higher hardness ratio (HR). It is common to use HR measurements as a proxy for the detection of obscuration events (\citealt{Kaa14}). In Fig.~\ref{hrfig} we show the soft and hard X-ray fluxes, and the hardness ratio measured for NGC 985 in different observations in the last two decades. The soft X-ray flux hit a minimum in 2013, and it increases again in 2015 by more than one order of magnitude, whereas the hard X-ray flux, less prone to be affected by absorption, has varied less than a factor of 2 or 3. Indeed, the HR of NGC 985 in 2013 was much higher than in 2015 (0.66 versus $\sim 0$), but the latter values are still slightly higher than the expected HR value in the case that no obscuration is taking place, which is of the order of $-0.14$. This value is the average HR from \swift{} observations in 2007-2008 when the source showed a persistently high soft X-ray flux, indicative that it was in an unobscured state. This would mean that in 2015 we are catching the source emerging from an absorbed state caused by intervening material but it is not yet fully unabsorbed. Furthermore, the HR in 2003 is also much higher than the unabsorbed value, also supported by the best-fit results shown in Table~\ref{xtab}. Since there were no hints of absorption in 2002, the source could be entering a similar absorbed state in 2003, which would mean that these eclipsing events are recurrent given that \swift{} monitoring between 2007 and 2008 indicated high soft X-ray flux levels (\citealt{Par14}).

We cannot determine a precise ionization parameter or total column density for the UV absorbing gas in the 2013 COS spectrum since we merely have a set of lower limits on ionic column densities, not definitive measurements. However, we can say that the range of ions we observe (\ion{C}{iii}, \ion{Si}{iv}, and \ion{C}{iv}) are consistent with gas having an ionization parameter in the range log $\xi \sim -1.0$ to $-0.5$. Our lower limits on the column densities are also compatible with the total column density seen in the low-ionization X-ray obscurer, {\it xabs} 4. For example, if we use log $\xi=-0.94$, which corresponds to X-ray absorbing component {\it xabs} 4 in the 2013 observation, our most stringent lower limit is for \ion{Si}{iv}, for which we get $\rm N_{\rm H} > 5 \times 10^{19}$~cm$^{-2}$. Therefore, it is likely that the same gas causing the low-energy X-ray obscuration is also the gas producing the broad UV absorption lines.

We also can not establish a precise location for the broad UV absorbing gas, but the covering fractions we discussed in Sect.~\ref{results} suggest that the gas covers more of the continuum source than the BLR. Therefore, the gas is likely in or near the BLR, which is consistent with the high velocities we observed both in width and in radial motion. 

The appearance of broad, high-velocity UV absorption at the same time as the strong, soft X-ray obscuration in NGC 985 strongly suggests that the obscurer is due to a wind associated with an accretion disk. This is now the third possible obscuration event detected in otherwise unobscured Seyfert 1 galaxies. The most thoroughly studied other example is the X-ray and UV obscurer that appeared in NGC 5548 (\citealt{Kaa14}), which still appears to be present. In addition, the weak \ion{C}{iv} absorption detected in Mrk 335 (\citealt{Lon13}) several months after a low-flux obscuration event is similar to the behavior we have observed in NGC 985, where we see weak UV absorption still present many months (18) after the original X-ray obscuration.

In NGC 5548, the X-ray and UV obscurer may cover as much as 100\% of the ionizing continuum (\citealt{Kaa14}); this shadowing results in a dramatic suppression of the ionizing flux that is manifested by the presence of many lower ionization absorption lines associated with the persistent, narrow UV absorption lines (\citealt{Arav15}). In NGC 985, since the UV obscuration appears not to be as deep and covering perhaps no more than 27\% of the continuum, the suppression of ionizing flux is not as dramatic. We do not see the same lowering of ionization state of the narrow UV absorbers as was observed in NGC 5548.

The sporadic nature of obscuration associated with high velocity outflows in Seyfert 1 galaxies is consistent with a picture in which the outflows are largely confined to the plane of the accretion disk, and only occasionally cross our line of sight as streamers or clumps rising chaotically from the edges of the outflow. In such a picture, broad absorption-line quasars, where such obscuration is the rule rather than the exception, would represent a class of objects viewed at higher inclination than typical Seyfert 1s. Continued monitoring of bright Seyfert 1s to trigger multiwavelength follow-up observations can yield greater insights into these phenomena.

\begin{acknowledgements}
This work is based on observations obtained by \XMM{}, an ESA science mission with instruments and contributions directly funded by ESA member states and the USA (NASA). SRON is supported financially by NWO, the Netherlands Organization for Scientific Research. This work was supported by NASA through a grant for HST program number 13812 from the Space Telescope Science Institute, which is operated by the Association of Universities for Research in Astronomy, Incorporated, under NASA contract NAS5-26555. We thank the anonymous referee for his/her useful comments that helped to improve the paper.
\end{acknowledgements}

\end{document}